\documentstyle[12pt]{article}
\begin{document}
\def\beq{\begin{equation}}
\def\eeq{\end{equation}}
 
\begin{flushright}
UMDGR-98-22\\
gr-qc/9709048\\
\end{flushright} 
\vskip 5mm

\begin{center}
 
{\Large Comment on\\
``Accelerated Detectors and Temperature\\ in (Anti) de Sitter Spaces"\\}
 
\vskip 5mm
{Ted Jacobson\footnote{E-mail: jacobson@physics.umd.edu}
\\Department of Physics, University of
Maryland\\ College Park, MD 20742-4111, USA\\} 
\end{center}
 
\vskip 5mm

\begin{abstract}
{It is shown how the results of Deser and Levin on the response of
accelerated detectors in anti-de Sitter space 
can be understood from the same general perspective as 
other thermality results in spacetimes
with bifurcate Killing horizons.}
\end{abstract}

\vskip 1cm
A detector with linear acceleration $a$
in the Minkowski vacuum sees a thermal bath
at the temperature $T_U=a/2\pi$ \cite{Unruh},
while an inertial detector in de Sitter (dS) space  
of radius $R$ sees a thermal bath
in the de Sitter vacuum at the temperature $T_{GH}= 1/2\pi R$
\cite{GH}.  
What does an accelerated detector see in 
de Sitter space?  
This detector
also sees a thermal bath, but at the temperature\cite{Pf,NPT,DL}
\begin{equation}
T_{dS}=(R^{-2} + a^2)^{1/2}/2\pi.
\label{TdS}
\end{equation}
Deser and Levin (DL) recently showed \cite{DL} that the 
same formula with $R^2\rightarrow -R^2$ 
gives in anti-de Sitter (adS) space the temperature seen 
by some uniformly accelerated detectors in any of three vacuum states,
while the temperature for some other uniformly accelerated
detectors vanishes! (In the adS
case the class of accelerated world lines yielding (\ref{TdS})
has acceleration bounded below by  $R^{-1}$ so the 
argument of the square root is bounded 
below by zero.) 

In the dS case DL obtain the result (\ref{TdS}) by viewing 
de Sitter space as a timelike hyperboloid embedded in five dimensional
Minkowski space, and invoking the fact \cite{Tagirov} 
that the Wightman function
for a conformally coupled massless scalar field in dS is just the induced
Wightman function from the 5d Minkowski vacuum. 
Since the uniformly accelerated worldlines in de Sitter are also
uniformly accelerated in the 5d Minkowski embedding space,
the detector
response is the same as in the 5d Minkowski vacuum, which appears as
a thermal state at the Unruh temperature $T=a_5/2\pi$. The result (\ref{TdS})
then follows from 
\begin{equation}
a_5=(R^{-2} + a^2)^{1/2}.
\label{a5}
\end{equation}

AdS space is the hyperboloid
$(z^0)^2-(z^1)^2-(z^2)^2-(z^3)^2+(z^4)^2=R^2$ 
in a flat 5d ``ultra hyperbolic" embedding space.
DL compute the detailed form of 
the transition rate in first order perturbation theory
for a detector on two classes of 
uniformly accelerated worldlines in adS, (i) circles 
$(z^0)^2 +(z^4)^2= \mbox{const}$ at fixed $(z^1,z^2,z^3)$, 
and (ii) hyperbolas $(z^0)^2-(z^1)^2 = \mbox{const} <0$ at
fixed $(z^2,z^3,z^4)$.
Since adS is not globally
hyperbolic there is no unique ``vacuum" state, but DL examine
three different choices for the adS quantum field state
corresponding to different boundary conditions at infinity \cite{AIS}.
For all of these states the detector transition rate vanishes
for worldlines of class (i), and for those of class (ii)
the rate has a Planck factor
at the temperature (\ref{TdS}), although the actual rate in two of the three
cases has an energy dependent prefactor.

The purpose of this comment is to indicate another 
way of arriving at some of the same 
conclusions which emphasizes the role of the
bifurcate Killing horizon, the global thermal nature of the states,
and the connection with related results in Minkowski, de Sitter,
and Schwarzschild spacetimes.
Let us begin with de Sitter space, which has no globally
timelike Killing field. A boost-like dS Killing
field vanishes on the equatorial $S^2$ of an $S^3$ spatial slice.
This $S^2$ is the bifurcation surface of the Killing horizon.
In terms of the Hamiltonian $H_B$ that generates
evolution along this Killing field, the density matrix of the 
de Sitter vacuum restricted to one side of the Killing horizon takes the 
form\cite{LaFlamme,TJ}
\begin{equation}
\rho=Z^{-1}\exp(-2\pi H_B).
\label{rho}
\end{equation}
In writing (\ref{rho}) the Killing field has been taken to be 
$\xi=\partial/\partial \theta$, where on the Euclidean section the range
of the dimensionless variable $\theta$ is $2\pi$. 
Thus the de Sitter vacuum is thermal at the dimensionless temperature
$T=1/2\pi$. To determine the temperature $T'$ 
seen by an observer on any
orbit of this Killing field
note that if $\xi$
has norm $N$ on this orbit, then the rescaled Killing field
$\xi'= N^{-1} \xi$ has unit norm there, so the Hamiltonian appropriate
to this observer scales as $H_B'= N^{-1} H_B$,
so that the effective temperature scales as 
\beq
T'=N^{-1} T=N^{-1}/2\pi.
\label{T'}
\eeq
This is just the Tolman redshift. 
It is easy to see that on a Killing orbit
of acceleration $a$ the norm $N$ of $\xi$ is just the inverse
of the ``five"-dimensional acceleration (\ref{a5}),
so (\ref{T'}) yields (\ref{TdS}).

In adS there {\it is} a globally timelike Killing field which generates
rotations in the $z^0z^4$ plane of the 5d embedding space.
This is $\tau$-translation in the intrinsic 4d line element (9) of \cite{DL}.
A detector following an orbit of this Killing field will remain
unexcited in the ground state of the Hamiltonian $H_\tau$ that
generates this symmetry. The three states considered by DL
are ground states for $H_\tau$ corresponding to different boundary
conditions at infinity, hence there is no detector excitation
for these orbits. 

There is also in adS
a boost-like Killing field with bifurcate Killing horizon.
I will now argue that the ground state of $H_\tau$ is a thermal
state relative to the Hamiltonian $H_B$ generating the boost-like
Killing symmetry.
On the covering space, with the timelike direction unwrapped, 
the bifurcation surface is an infinite collection of hyperbolic planes.
In spite of the lack of global hyperbolicity, the usual Euclidean path
integral construction\cite{LaFlamme} of the  
wavefunctional for the ground state of $H_\tau$ 
on a spatial slice through one of the bifurcation surfaces
can still be used. 
I don't know of anywhere this path integral is carried
out explicitly for the adS case, but it seems clear that  
it depends upon the boundary conditions imposed at infinity
and can yield any of the states considered in \cite{AIS}.
As in Minkowski, de Sitter, and Schwarzschild
spacetimes, slicing this path integral in ``polar" coordinates
adapted to the Euclideanized boost-like symmetry reveals that,
restricted to one side of the bifurcation surface,
the state is a thermal density matrix of the form (\ref{rho}). 

What temperature do different boost-like Killing observers
see in adS?  The norm of the Killing field
on a Killing  orbit of acceleration $a$ can be obtained from 
the dS case
by analytic continuation of $R^2$ to negative values, 
so (\ref{T'}) yields 
the result of DL, 
\beq
T_{adS}=(-R^{-2} + a^2)^{1/2}/2\pi,
\label{TadS}
\eeq
which is just (\ref{TdS}) with $R^2\rightarrow -R^2$. 
The acceleration of the adS Killing orbits approaches the constant
$R^{-1}$ at infinity (unlike in Rindler or Schwarzschild spacetime
where it vanishes). Thus $T_{adS}$ (\ref{TadS}) vanishes at infinity,
even though the Killing orbits are still accelerating there.
This is because, as (\ref{T'}) shows, it is not the
acceleration but the norm of the Killing field that
determines the locally observed temperature in the
state defined by the Euclidean path integral. 
The norm of the relevant adS Killing field
diverges at infinity, so the observed temperature vanishes
there, as in the Minkowski vacuum in Rindler space.
In Schwarzschild spacetime, conversely, the 
norm of the Killing field is finite at infinity, so the
local temperature is nonzero there even though the
acceleration vanishes. This is just
the Hawking temperature of the Hartle-Hawking state at infinity. 

I have argued that the ground state of $H_\tau$ is a thermal state
relative to the Hamiltonian $H_B$ generating the boost-like Killing 
symmetry. This is a general conclusion which applies (at least
formally) to {\it any} field theory, including interacting ones.
Nevertheless,
detectors will respond differently in the ground states corresponding
to different field theories. Even in free field theory,  
boundary conditions will influence the detector response due to 
the different nature of the modes that generate the Hilbert space.
The calculation of Deser and Levin 
reveals precisely how the different boundary
conditions affect the detector response for a conformally coupled
massless free field, which is something that cannot
be obtained from (\ref{rho}) without an explicit construction of
the operator $H_B$ in each case.  

\vskip 5mm
I am grateful to D. Brill, S. Deser, O. Levin and J. Louko 
for helpful conversations.
This work was supported in part by NSF grant PHY94-13253.


\begin{thebibliography}{99}
\bibitem{Unruh}Unruh W.G. 1976 Phys. Rev. D {\bf 14} 870 
\bibitem{GH}Gibbons, G.W. and Hawking, S.W. 1977 Phys Rev. D {\bf 15} 2738
\bibitem{Pf}Pfautsch, J., cited in Davies, P.C.W. 
1984 Phys. Rev. D {\bf 30} 737
\bibitem{NPT}Narnhofer, H., Peter, I. and Thirring, W. 1996 Int. J. Mod. Phys.
B {\bf 10} 1507
\bibitem{DL}Deser, S. and Levin, O. 1997 Class. Quantum Grav. {\bf 14} L163
[gr-qc/9706018]
\bibitem{Tagirov}Tagirov, E.A. 1973 Ann. Phys. {\bf 76} 561
\bibitem{AIS}Avis, S.J., Isham, C.J. and Storey, D. 1978 Phys. Rev. D {\bf 18}
3565
\bibitem{LaFlamme}LaFlamme, R. 1989 Nucl. Phys. {\bf B324} 23
\bibitem{TJ}Jacobson, T. 1994 Phys. Rev. D {\bf 50} R6031 [gr-qc/9407022]

\end{thebibliography}
\end{document}